\documentstyle[12pt]{article}
\def\be {\begin{equation}}
\def\ee {\end{equation}}
\def\ba {\begin{eqnarray}}
\def\ea {\end{eqnarray}}
\newcommand{\bq}{\begin{eqnarray}}
\newcommand{\eq}{\end{eqnarray}}

\def\bi {\begin{itemize}}
\def\ei {\end{itemize}}
\begin{document}
\def\bea{\begin{eqnarray}}
\def\eea{\end{eqnarray}}
\title{\bf {Polytropic inspired inflation }}
\author{M. R. Setare\footnote{rezakord@ipm.ir}\\ 
\emph{Department of Science,University of Kurdistan, Sanandaj, Iran,
}\\
{F. Darabi\footnote{f.darabi@azaruniv.edu (Corresponding author)}}\\
\emph{Department of Physics, Azarbaijan Shahid Madani University, Tabriz, Iran. }}

\date{\small{}}
\maketitle
\begin{abstract}
We study the chaotic inflation in the context of a gravity theory where 
the Friedman equation is modified, inspired by the polytropic gas equation of state. It is seen that in the $n=1$ case for the polytropic index the inflaton field at the end of inflation $\phi_e$, depends on the Planck mass, while for $n\neq1$ it generally depends on the polytropic constant and mass of the inflaton field.
\end{abstract}
\\
Pacs: 98.80.Cq

\section{Introduction}

The current accelerated expansion of the universe has now become an inevitable
fact in the context of cosmology. This fact is based on the observations which are obtained by SNe Ia \cite{c1}, WMAP \cite{c2}, SDSS \cite{c3}
and X-ray \cite{c4}. These observations also suggest that our
universe is spatially flat, and consists of about $70 \%$ dark
energy with negative pressure, $30\%$ dust matter (cold dark
matter plus baryons), and negligible radiation. In the framework of standard cosmology, a missing energy component with negative pressure so called dark energy (DE) is responsible for this expansion. The simplest candidate of the DE is a tiny positive time-independent cosmological constant. However, the cosmological constant suffers from two well known ``fine-tuning" and ``cosmic coincidence" problems. The first problem is that: while we may choose the value of the cosmological constant to fit with the current observations, the theoretical estimates of this value remain about 120 orders of magnitude too large. The second problem is that: why are we living in an epoch in which the energy densities of cosmological constant and dust matter are comparable?
Facing with these puzzles, it is not surprising that the search for alternative forms of DE is an ongoing one. The dynamical nature of dark
energy can originate from various fields. There are two categories for dynamical
DE: (i) The scalar fields including: a canonical scalar field called quintessence \cite{quint1,quint2,quint3}, a phantom field with a negative sign of the kinetic term \cite{phant1,phant2,phant3,phant4} , the combination of quintessence and phantom in a unified model called quintom
\cite{quintum}, and so forth, (ii) The interacting DE models including:
Chaplygin gas models with a perfect fluid having a novel equation of state
(EoS) \cite{Chaplyg}, holographic models constructed in the light of the holographic principle of quantum gravity theory \cite{holograph}.

Among the above list, the Generalized Chaplygin Gas (GCG) is an
interesting model of dynamical DE which can mimic the
behavior of matter at early-time universe and that of a cosmological
constant at late-time universe \cite{GCG}. Furthermore, the
inflation can be accommodated within the GCG scenario in a Chaplygin
inspired inflation model \cite{Bertol}. In this model, the GCG background is described by an exotic EoS
\begin{equation}\label{-}
p_{GCG}=-\frac{A}{\rho_{GCG}^{\alpha}},
\end{equation}
and energy density
\begin{equation}\label{0}
{\rho}_{GCG}=\left[A+\frac{B}{a^{3(1+\alpha)}}\right]^{\frac{1}{1+\alpha}},
\end{equation}
where $A$, $B$, and ${\alpha}$ are positive constants and $a(t)$ is the scale
factor of the universe. The case ${\alpha=1}$ corresponds to the Chaplygin gas EoS. The equation (\ref{0}) indicates that, as the universe evolves, the GCG energy density interpolates between the energy density of non-relativistic matter (second term) and that of a cosmological constant (first term). This interesting property allows one to interpret the GCG as DE with an admixture of dark matter. 

The authors in \cite{Bertol}, however, did not view Eq.(\ref{0}) as a consequence of the EoS (\ref{-}), rather, they assumed it arising due to a modification of gravity through a modified Friedmann equation of the form
\begin{equation}\label{0'}
H^2=\frac{1}{3M^2}\left[A+\rho_{\phi}^{(1+\alpha)}\right]^{\frac{1}{1+\alpha}},
\end{equation}
where $\rho_{\phi}$ is the energy density of the inflaton field. This modification is Chaplygin inspired because it follows from an extrapolation of Eq.(\ref{0})
\begin{equation}\label{0''}
{\rho}_{GCG}=\sqrt{A+\rho_m^2} \:\:\: \longrightarrow \:\:\: \sqrt{A+\rho_{\phi}^2} \end{equation}
where ${\rho}_{m}$ corresponds to the matter energy density (we
set $\alpha = 1$ for simplicity). The proposal that GCG model may be viewed as a modification of gravity was first pointed out in \cite{Barr}.

On the other hand, the
polytropic gas has been proposed as an alternative model for
describing the accelerating of the universe \cite{12}. Polytropic
EoS has been used in various astrophysical situations,
for example it can explain the EoS of degenerate white
dwarfs, neutron stars and also the EoS of main
sequence stars \cite{13}, and in the case of Lane-Emden models
\cite{14,15}. Recently, people have also investigated the cosmological implications of polytropic gas DE model, polytropic gas model from the viewpoint of statefinder analysis and Polytropic scalar field models of dark energy \cite{26,27,28}.  
As mentioned previously chaotic inflation in the
context a phenomenological modification of gravity inspired by the
Chaplygin gas EoS has been studied by Bertolami et al
\cite{Bertol}. According to this model, the scalar field, which
drives inflation, is the standard inflaton field, can be extrapolate
for obtaining a successful inflation period with a Chaplygin gas
model. After that this scenario extended to the Chaplygin inspired
inflationary model in which a brane-world model is considered
\cite{17}.

Based on the above statements we are motivated to consider early
universe cosmological implications of this model and investigate if
it can inspire inflation like the model of {\it Chaplygin Inspired
Inflation} \cite{Bertol}.

\section{Polytropic gas as an effective inflaton field}

The EoS in polytropic gas model is given by \cite{13}
\begin{equation}\label{1}
p_{\Lambda}=K\rho_{\Lambda}^{1+\frac{1}{n}},
\end{equation}
where $p_{\Lambda},~ \rho_{\Lambda}$, $K$ and $n$ are the pressure, energy density, polytropic constant and polytropic index, respectively. The conservation equation in
Friedmann-Robertson-Walker universe is given by
\begin{equation}\label{2}
\dot{\rho}_{\Lambda}+3H(\rho_{\Lambda}+p_{\Lambda})=0,
\end{equation}
where $H$ is the Hubble parameter and a dot denotes differentiation with
respect to the cosmological time. Substituting the EoS (\ref{1}) into the conservation equation, we obtain
\begin{equation}\label{3}
\rho_{\Lambda}=[-K+Ba^{\frac{3}{n}}]^{-n},
\end{equation}
where $a(t)$ is the scale factor of the universe and $B$ is a
positive integration constant. As in Ref.\cite{GCG}, this result indicates
that the polytropic energy density interpolates between that of non-relativistic matter $Ba^{-3}$ and that of a cosmological constant $-K^{-n}$, so one may
interpret the polytropic energy density as an admixture of DE with
dark matter. For $n=1$, we may use of the following extrapolation to an inflationary era
$$
\rho_{\Lambda}=[-K+\rho_{m}^{-1}]^{-1} \longrightarrow \rho_{\Lambda}=[-K+\rho_{\phi}^{-1}]^{-1},
$$
where $\rho_{m}$ is the matter energy density and $\rho_{\phi}$ is
the energy density of an effective inflaton field $\phi$. This provides us with a modified Friedmann equation for arbitrary $n$ as
\begin{equation}\label{4}
H^2=\frac{1}{3M_P^2}[-K+\rho_{\phi}^{-\frac{1}{n}}]^{-n},
\end{equation}
where ${M_P}$ is the Planck mass, and it is assumed that dynamics of the
the inflaton field is given by
\begin{equation}\label{5}
\ddot{\phi}+3H\dot{\phi}+V^{\prime}(\phi)=0,
\end{equation}
where $V(\phi)$ is the inflaton field potential and $'$ denotes differentiation
with respect to $\phi$.

\section{Slow roll approximations }

The energy density and pressure of the inflaton field is given by
\begin{equation}\label{6}
\rho_{\phi}=\frac{1}{2}\dot{\phi}^2+V(\phi).
\end{equation}
\begin{equation}\label{66}
p_{\phi}=\frac{1}{2}\dot{\phi}^2-V(\phi).
\end{equation}
Assuming the slow roll approximations
$$
\dot{\phi}^2\ll V(\phi),~~~~~\ddot{\phi}\ll V^{\prime}(\phi),
$$
we obtain
\begin{equation}\label{7}
H^2\simeq \frac{1}{3M_P^2}[-K+[V(\phi)]^{-\frac{1}{n}}]^{-n},
\end{equation}
\begin{equation}\label{8}
\dot{\phi}\simeq -\frac{\sqrt{3}}{3}M_P V^{\prime}(\phi)[-K+[V(\phi)]^{-\frac{1}{n}}]^{n/2},
\end{equation}
where use has been made of $\rho_{\phi}\simeq V(\phi)$. Differentiation of Eqs.(\ref{4}), (\ref{8}) with respect to time and using $\rho_{\phi}\simeq V(\phi)$ and $\rho_{\phi}+p_{\phi}=\dot{\phi}^2$ leads respectively to
\begin{equation}\label{9}
\dot{H}\simeq -\frac{1}{6}[V^{\prime}(\phi)]^2[V(\phi)]^{-\frac{1}{n} -1} [-K+[V(\phi)]^{-\frac{1}{n}}]^{-1},
\end{equation}
\begin{equation}\label{10}
\ddot{\phi}\simeq -\frac{\sqrt{3}}{3}M_P \dot{\phi}[-K+[V(\phi)]^{-\frac{1}{n}}]^{\frac{n}{2}}
\left[V^{\prime\prime}({\phi})-\frac{1}{2}[V^{\prime}(\phi)]^2[V(\phi)]^{-\frac{1}{n}-1}[-K+[V(\phi)]^{-1}\right].
\end{equation}
The slow roll parameters are then obtained as follows
\begin{equation}\label{11}
\epsilon \equiv -\frac{\dot{H}}{H^2}\simeq \frac{[M_P V^{\prime}(\phi)]^2}{2}
[V(\phi)]^{-\frac{1}{n} -1} [-K+[V(\phi)]^{-\frac{1}{n}}]^{n-1}\ll 1,
\end{equation}
\begin{equation}\label{12}
\delta \equiv -\frac{\ddot{\phi}}{H\dot{\phi}}\simeq {[M_P^2 V^{\prime\prime}(\phi)]}
[-K+[V(\phi)]^{-\frac{1}{n}}]^{n}-\epsilon\ll 1,
\end{equation}
\begin{equation}\label{13}
\eta\equiv\epsilon+\delta = \simeq {[M_P^2 V^{\prime\prime}(\phi)]}
[-K+[V(\phi)]^{-\frac{1}{n}}]^{n}\ll 1+\epsilon \simeq1.
\end{equation}
The condition $\epsilon \ll 1$ is evidently equal to the slow roll
condition $\dot{\phi}^2\ll V(\phi)$ which is necessary for the inflation. However, we have
\begin{equation}\label{14}
\frac{\dot{\phi}^2}{V(\phi)}=\frac{2}{3}[1-K[V(\phi)]^{\frac{1}{n}}]\epsilon,
\end{equation}
which is small through $\epsilon \ll 1$ if and only if $K[V(\phi)]^{1/n}\leq 1$. In fact, this condition should be valid throughout the polytropic regime signaling the onset of the polytropic inflation. Now, we demand $\phi_e$ to be the value of the inflaton field amplitude when inflation ends. This amplitude can be obtained by letting $\epsilon \leq 1$ which accounts
for the end of inflation, and casts into the following form
\begin{equation}\label{15}
\frac{{[V'(\phi)]}^4}{[V(\phi)]^{2(1+\frac{1}{n})}}\leq \frac{4}{M_P^4}
[-K+[V(\phi)]^{-\frac{1}{n}}]^{-2(n-1)}.
\end{equation}
Taking the chaotic inflation case $V(\phi)=m^2\phi^2/2$ we obtain the approximate equation for $\phi_e$ as
\begin{equation}\label{16}
\frac{m^{4(1+\frac{1}{n})}}{4^{\frac{1}{n}}M_P^{4}}\phi_e^{4(1+\frac{1}{n})}[-K+[m^2\phi_e^2/2]^{-\frac{1}{n}}]^{-2(n-1)}-m^8 \phi_e^4\simeq0.
\end{equation}
Solving this equation for arbitrary $n$ may be complicated, so for simplicity,
we will find the solutions for the following special cases.\\
For $n=1$ the equation takes on the simple form
\begin{equation}\label{17}
m^{8}\phi_e^{4}\left(\frac{\phi_e^4}{4M_P^{4}}-1\right)\simeq0,
\end{equation}
whose nontrivial solution is obtained as
\begin{equation}\label{18}
\phi_e\simeq \sqrt{2}M_P.
\end{equation}
For $n=2$ the equation becomes
\begin{equation}\label{19}
\frac{2}{m^2}\phi_e^{-4}-\frac{2\sqrt{2}K}{m}\phi_e^{-3}+K^2\phi_e^{-2}-\frac{1}{2M_P^4m^2}\simeq0,
\end{equation}
which has the nontrivial solutions
\begin{eqnarray}\label{20}
\phi_e\simeq\left(-\frac{\sqrt{2}}{2}KM_P m\pm\frac{1}{2}\sqrt{2(KM_P m)^2+8}\right)M_P,\\
\nonumber
\phi_e\simeq\left(+\frac{\sqrt{2}}{2}KM_P m\pm\frac{1}{2}\sqrt{2(KM_P m)^2+8}\right)M_P.
\end{eqnarray}
It is seen that except for the case $n=1$ for which $\phi_e$ depends just on the Planck mass, the solutions $\phi_e$ for $n\neq1$
generally depend on the polytropic constant and mass of the scalar field.
In other words, in the case of $n=1$, the duration of inflation is merely determined by the Planck mass, while those of $n\neq1$ are determined by
the polytropic constant and mass of the scalar field.

The number of e-folding during inflation is given by
\begin{equation}\label{21}
N(\phi_b \rightarrow \phi_e)\simeq \int_{\phi_e}^{\phi_b}~d\phi\frac{V(\phi)}{V^{\prime}(\phi)}\simeq
\int_{\phi_e}^{\phi_b}~d\phi\frac{1}{M_P^2V^{\prime}(\phi)}[-K+[V(\phi)]^{-\frac{1}{n}}]^{-n},
\end{equation}
where $\phi_b$ is the field amplitude when the inflation begins. For simplicity,
we evaluate the integral for $n=1$. Substituting the quadratic potential
$V(\phi)=m^2\phi^2/2$ in the integrand, we obtain
\begin{equation}\label{22}
N(\phi_b \rightarrow \phi_e)\simeq
\int_{\phi_e}^{\phi_b}~d\phi\frac{1}{M_P^2m^2\phi}\left[-K+\left[\frac{2}{m^2\phi^2}\right]\right]^{-1}
=\frac{1}{2Km^2M_P^2} \ln\left[\frac{2-Km^2\phi_e^2}{2-Km^2\phi_b^2}\right].
\end{equation}
We note that the condition $K[V(\phi)]^{1/n}\leq 1$, for $n=1$, reads
as $m^2\phi^2\leq 2/K$ which according to (\ref{22}) is valid throughout the polytropic regime producing the polytropic inflation.
If we substitute $\phi_e\simeq \sqrt{2} M_P$ into the above expression and solve  for $\phi_b$ in terms of $N$ we obtain
\begin{equation}\label{23}
\phi_b\simeq\left[\frac{2}{Km^2}\left[1-(1-Km^2M_P^2)e^{-2K N m^2M_P^2}\right] \right]^{\frac{1}{2}}.
\end{equation}
By expanding the exponential term we find
\begin{equation}\label{24}
\phi_b\simeq \sqrt{2(1+2N)}M_P,
\end{equation}
which asserts that a sufficient amount of inflation (Large $N$) can occur for values of $\phi_b$ more larger than $\phi_e$.

\section{Observational bounds}

In this section, we investigate polytropic inflation versus CMBR data. In this regard, we shall compute the amplitude of scalar perturbations \cite{Bertol,Lidsey}
\begin{equation}\label{24'}
A_S^2=\frac{4}{25}\langle\zeta^2\rangle,
\end{equation}
where $\zeta$ is the gauge invariant quantity
\begin{equation}\label{25'}
\zeta=\psi+H\frac{\delta \rho}{\dot{\rho}},
\end{equation}
$\psi$ and $\delta\rho$ being the gravitational potential and density fluctuation.
On the slices of uniform density, $\zeta$ reduces to the curvature perturbation
and on super-horizon scales the curvature perturbation is equal to the comoving curvature perturbation. Hence, in a spatially flat gauge, we have \cite{Bertol}
\begin{equation}\label{26'}
\zeta=H\frac{\delta \rho}{\dot{\rho}}=H\frac{\delta \phi}{\dot{\phi}},
\end{equation}
where $|\delta\phi|=H/2\pi$, is the freezed value on super-horizon scales at the time of horizon crossing. Using Eqs.(\ref{7}) and (\ref{8}) for $n=1$, we obtain $\zeta$ and hence the amplitude of scalar perturbations is given by
\begin{equation}\label{27'}
A_S^2=\frac{1}{75\pi^2 \beta^2M_P^2 m^4}\left(-KM_P^2+\frac{2}{m^2 \beta^2}\right)^{-3},
\end{equation}
where $\beta=\langle\phi_b\rangle$. Introducing the COBE's observational upper bound $A_S = 2\times10^{-5}$, results in
\begin{equation}\label{28'}
\frac{1}{75\pi^2 \beta^2 M_P^2 m^4}\left(-KM_P^2+\frac{2}{m^2 \beta^2}\right)^{-3}<4\times
10^{-10}.
\end{equation}
If we define $x = m^2/M_P^2$, $y =KM_P^4$, and put $\beta=14.8$, then the required upper bounds on the polytropic constant and mass of the inflaton field may be obtained by the following inequality
\begin{equation}\label{29'}
x^{-2}\left(-y+\frac{2}{\beta^2x}\right)^{-3}<3\beta^2\times
10^{-7},
\end{equation}
for which all observational constraints can be satisfied.

\section{Conclusion}

In this paper we have studied the chaotic inflation in the context a modified gravity model inspired
 by the polytropic gas EoS. The important
results we obtained are as follows: Chaotic inflation is possible
in the context of a phenomenological modification of gravity inspired
by the polytropic gas EoS. The $n=1$ case for the polytropic index gives an end-inflaton
field which merely depends on the Planck mass, while $n\neq1$ cases give end-inflaton fields which generally depend on the polytropic constant and mass of the inflaton field. In other words, in the case of $n=1$, the duration of inflation is merely determined by the Planck mass, while that of $n\neq1$ cases is determined by the polytropic constant and mass of the scalar field.

A comparison with the previous results obtained in the Chaplygin inspired
inflation model shows, in principle, that in the present model as well as Chaplygin inspired model, a sufficient amount of inflation can occur satisfying all observational constraints.

\end{document}